\documentclass{llncs}
\usepackage{llncsdoc}
\usepackage{graphicx}
\usepackage{hyperref}

\begin{document}
\savebox{1}{\url{https://doi.org/10.1007/978-3-030-86976-2_15}}

\title{
Object-oriented implementation of algebraic multi-grid solver
for lattice QCD on SIMD architectures and GPU clusters\thanks{
The final authenticated publication is available online at\\ \usebox{1}.
}}

\author{Issaku Kanamori\inst{1}
   \and Ken-Ichi Ishikawa\inst{2}
   \and Hideo Matsufuru\inst{3}
}

\institute{
 RIKEN Center for Computational Science (R-CCS), \\
 7-1-26, Minatojima Minamimachi, Kobe 650-0047, Japan\\
 \email{kanamori-i@riken.jp}
\and 
Core of Research for the Energetic Universe,\\
Graduate School of Advanced Science and Engineering, Hiroshima University,\\
1-3-1 Kagamiyama, Higashi-Hiroshima 739-8526, Japan 
 \email{ishikawa@theo.phys.sci.hiroshima-u.ac.jp}
\and 
 High Energy Accelerator Research Organization (KEK),\\
 1-1 Oho, Tsukuba 305-0801, Japan\\
 \email{hideo.matsufuru@kek.jp}
}

\maketitle

\begin{picture}(0,0)(0,0)
 \put(270,250){\makebox{HUPD-2111}}
\end{picture}

\begin{abstract}
A portable implementation of elaborated algorithm is important to 
use variety of architectures in HPC applications.
In this work we implement and benchmark an algebraic multi-grid solver for Lattice QCD on three different architectures,
Intel Xeon Phi, Fujitsu A64FX, and NVIDIA Tesla V100,
in keeping high performance and portability of the code based on
the object-oriented paradigm. 
Some parts of code are specific to an architecture employing appropriate
data layout and tuned matrix-vector multiplication kernels, while
the implementation of abstract solver algorithm is common to all architectures.
Although the performance of the solver depends on tuning
of the architecture-dependent part, we observe reasonable scaling behavior and
better performance than the mixed precision BiCGSstab solvers.

\end{abstract}

\section{Introduction}
\label{sec:Introduction}

Lattice QCD simulations have been one of the most challenging
subjects in high performance computing in science.
As increasing the precision of experimental data in elementary
particle and nuclear physics, theoretical calculation is
required to provide correspondingly precise predictions.
Lattice QCD simulations are not only the first principle
calculation of the strong interaction but also providing
nonperturbative procedure to examine the candidates of new
physics.
A typical bottleneck in numerical simulations of lattice
QCD is the linear equation for the Dirac fermion operator
that is a large sparse matrix.
As the simulated system becomes closer to the real system
the linear equation becomes more difficult to solve
because of larger lattice size and smaller quark mass parameter
that increases the condition number.
Therefore variety of improved algorithms have been developed
for solving this linear equation.
At the same time, numerous efforts have been devoted to
exploit the new architecture of computers.

Multi-grid algorithms are widely used in solving large linear system.
This type of algorithms applies coarsening of the lattice
to define a matrix of smaller rank which reflects the long
range effect of the original matrix while is easier to solve.
The solution on the coarse lattice is used as a preconditioner
of the original linear equation solver.
A simple geometric coarsening, however, does not accelerate
linear equation solvers in lattice QCD.
It has been demonstrated that algebraic coarsening with adaptive setup
is needed
\cite{Brannick:2007ue,Babich:2010qb,Osborn:2010mb,%
Cohen:2012sh,Frommer:2013fsa}.

A direct trigger of this work is a need of porting
the multi-grid algorithm to the supercomputer Fugaku~\cite{Fugaku} that
has been installed at RIKEN Center for Computational Science (R-CCS)
and provided for shared use recently.
To make use of its potential arithmetic performance,
one needs to implement a code that exploit the specific
structure of the architecture.
The A64FX architecture of Fugaku adopts the SIMD arithmetic
operation in units of 512-bit length.
While the same SIMD width is also adopted in
the Intel AVX-512 instruction set architecture, their
micro architectures are different, especially,
in the instructions across the SIMD lanes needed for 
complex number arithmetics.
Indeed different data layouts achieve better efficiency for
the AVX-512 and A64FX architectures, as shown later.

Another important target of this work is GPU clusters,
or systems with accelerator devices in general, which have
been increasingly adopted as recent large-scale supercomputers.
Such systems require heterogeneous parallelization code
for the host processors and many-core devices.
To implement the device kernel, several frameworks are
available such as OpenACC, CUDA, and OpenCL.
We adopt OpenACC in this work.

Under this situation, it is desirable to establish a code design
that keeps machine specific implementation minimum while allows
construction of algorithms in generic manner.
The goal of this paper is to develop such a framework to
implement the multi-grid algorithm applicable to various
architectures exemplified for A64FX, Intel AVX-512,
and an NVIDIA GPU cluster.
To realize these contradicting issues, we adopt
the object-oriented
program design and specify the interface that the architecture
specific code must provide.
As a framework for implementation, we adopt a general purpose
lattice QCD code set Bridge++ with extension to add a code that
is tuned for specific architecture.
We discuss how much the algorithm and optimization of code
for specific architecture can be separated.

This paper is organized as follows.
In the next section, we briefly introduce the linear equation
in lattice QCD that is the target problem of this work.
Section~\ref{sec:Algorithm} summarizes the multi-grid algorithm
implemented in this work.
Section~\ref{sec:Implementation} describes our code
implementation.
Section~\ref{sec:Performance} shows performance results
measured on the following systems: Oakforest-PACS of
JCAHPC (Intel Xeon Phi KNL cluster), Fugaku,
and the Cygnus system at Univ. of Tsukuba (NVIDIA V100 cluster).
The last section is devoted to conclusion and outlook.

\section{Linear equation in lattice QCD simulations}
\label{sec:LatticeQCD}

In this section, we introduce the linear equation problem
in lattice QCD simulations to least extent for understanding
the following sections.
For the formulation of lattice QCD and the principle of
numerical simulation, there are many textbooks and reviews
(e.g. \cite{Knechtli:2017sna}).

The lattice QCD is a field theory formulated on a four-dimensional
Euclidean lattice.
It consists of fermion (quark) fields and a gauge (gluon) field.
The latter mediates interaction among quarks and are represented by
`link variable', $U_\mu(x)\in SU(3)$, where $x=(x_1,x_2,x_3,x_4)$
stands for a lattice site and $\mu =1,2,3,4$ is the spacetime direction.
In numerical simulations the lattice size is finite:
$x_\mu =1,2,\cdots,L_\mu$.
The fermion field is represented as a complex vector on lattice sites,
which carries 3 components of `color' and 4 components of `spinor',
thus in total 12, degrees of freedom on each site.
The dynamics of fermion is governed by a functional
$S_F = \sum_{x,y} \psi^\dag(y) D[U]^{-1}(x,y) \psi(y)$,
where $D[U]$ is a fermion matrix acting on a fermion vector $\psi(x)$.
A Monte Carlo algorithm is applied to generate an ensemble of
the gauge field $\{ U_\mu (x) \}$, that requires to solve a linear
equation $x = D^{-1}\psi$ many times.

There is a variety of the fermion matrix $D[U]$ depending on
the way to discretize the continuum theory.
As a common feature, the matrix is sparse because of the locality
of the interaction among quarks and gluons.
In this paper, we examine the $O(a)$-improved Wilson fermion,
also called clover fermion matrix,
\begin{equation}
D_{x,y} = [1 + F(x)] \delta_{x,y} - \kappa \sum_{\mu=1}^4 
  \left[ (1-\gamma_\mu) U_\mu(x) \delta_{x+\hat{\mu}, y}
       + (1+\gamma_\mu) U^\dag_\mu (x-\hat{\mu}) \delta_{x-\hat{\mu},y} \right] ,
\label{eq:fermion_matrix}
\end{equation}
where $x$, $y$ are lattice sites, $\hat{\mu}$ the unit vector
along $\mu$-th axis, and the hopping parameter
$\kappa = 1/(8+2m_0)$ related to the quark mass $m_0$.
$F(x)$ is a $12\times 12$ Hermitian matrix made of link variables,
which is introduced to reduce  the finite lattice spacing artifact.
As mentioned above, the link variable $U_\mu(x)$ is a $3\times 3$
complex matrix acting on the color and $\gamma_\mu$ is a $4 \times 4$
matrix acting on the spinor degrees of freedom.
Thus $D$ is a complex matrix of the rank $4\cdot 3  L_x L_y L_z L_t$.
It is standard to impose the periodic or anti-periodic boundary conditions.

As a general feature, as decreasing the quark mass $m_0$, the linear
equation becomes more and more difficult to solve since the condition
number of the matrix increases.
Another general feature common to fermion matrices is so-called
$\gamma_5$-Hermiticity, which is a remnant of the Hermiticity of
fermion operator in the continuum theory:
\begin{equation}
D^\dag = \gamma_5 D \gamma_5 , \hspace{0.5cm}
\gamma_5 = - \gamma_1 \gamma_2 \gamma_3 \gamma_4 .
\label{eq:gamma5_hermiticity}
\end{equation}
The byte-per-flop of the clover fermion is 0.94 for single
precision arithmetics.

\section{Multi-grid algorithm}
\label{sec:Algorithm}

This section describes the multi-grid algorithm applied
in this paper.
We first settle the convention to represent vectors.
Adopting the style in the quantum mechanics,
we denote a vector as $| x \rangle$ and its Hermitian
conjugate $\langle x|$.
They are vectors in the original fine lattice, and
the vectors on coarse lattice is represented with
suffix `c', such as $|x\rangle_c$.
For simplicity we describe a single level multi-grid algorithm,
while a multi-level implementation is straightforward.
The linear equation to be solved is
\begin{equation}
  D |x \rangle = |b \rangle .
\label{eq:linear_equation_on_fine_lattice}
\end{equation}
In the following, we first describe the general feature of
the multi-grid algorithm and then technical details
specific to the lattice QCD problem.

\paragraph{Solver algorithm with preconditioner.}
The multi-grid algorithm works as a preconditioner of
iterative Krylov subspace solvers.
While there is a variety of choice for this outer solver,
in this work we employ the BiCGStab algorithm with flexible
preconditioner summarized as follows \cite{Vogel2007}.
\bigskip\\
\begin{tabular}{l}
\hline
\noalign{\smallskip}
BiCGStab algorithm with flexible preconditioning\\
\noalign{\smallskip}
\hline
\noalign{\smallskip}
$ |x\rangle := |x_0\rangle, \hspace{0.2cm}  |r\rangle := |b\rangle - D|x_0\rangle$, $|\tilde{r}\rangle := |r\rangle$, $|p\rangle = |r\rangle$  \\
$M$ is the preconditioner and $|x_0\rangle$ is the given initial guess. \\
for $i$ = 0, \dots, $n$ \ do \\
\hspace{0.5cm} $\alpha_i=\langle \tilde{r}|r_i\rangle / \langle \tilde{r}|D M |p_i\rangle$ \\
\hspace{0.5cm} $|x_i\rangle := |x_i\rangle +\alpha_i M|p_i\rangle$,\quad $|s_i\rangle := |r_i\rangle -\alpha_i DM|p_i\rangle$\\
\hspace{0.5cm} if($\bigl| |s_i\rangle\bigr|^2$ is small enough) break\\
\hspace{0.5cm} $|t_i \rangle := DM |s_i\rangle$, \quad 
$\omega_i := \langle t_i | s_i \rangle/ \langle t_i | t_i\rangle$ \\
\hspace{0.5cm} $|x_{i+1}\rangle := |x_i\rangle +  \omega_i M |r_i\rangle$, \quad $|r_{i+1}\rangle := |s_i\rangle -\omega_i DM |s_i\rangle$ \\
\hspace{0.5cm} if($\bigl| |r_{i+1}\rangle\bigr|^2$ is small enough) break \\
\hspace{0.5cm} $\beta_i=(\alpha_i/\omega_i) \langle \tilde{r}|r_{i+1}\rangle/\langle \tilde{r}|r_i\rangle$, \quad
$|p_{i+1}\rangle := |r_{i+1}\rangle - \beta_i(|p_i\rangle -\omega_i D |p_i\rangle$ \\
end \\
\noalign{\smallskip}
\hline
\end{tabular}
\bigskip

If the preconditioner $M$ provides an approximate solution,
$Mx \sim D^{-1}x$, the algorithm works efficiently.
As an example of such a preconditioner, multi-precision solver
is known to work efficiently, in particular on a system whose
performance is regulated by its memory bandwidth.
The version of algorithm we use has two break points as
described above.

\begin{figure}[t]
\centering
\includegraphics[width=8.0cm]{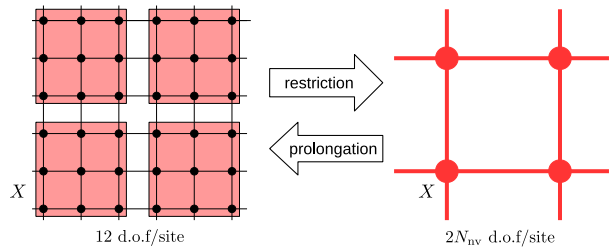}
\caption{
Schematic description of the multi-grid algorithm. A domain $X$ on the fine lattice is mapped to a site $X$ on the coarse lattice.
}
\label{fig:multi-grid_operations}
\end{figure}

\paragraph{Multi-grid algorithm.}
Figure~\ref{fig:multi-grid_operations} schematically explains
the idea of the multi-grid algorithm.
In some way we define the vectors on a coarser lattice,
$|b\rangle \rightarrow |b \rangle_c$.
This process is called `restriction'.
Defining the matrix $D_c$ multiplied to this coarse vector
correspondingly, one solves the linear equation
\begin{equation}
D_c |x \rangle_c  = |b\rangle_c .
\label{eq:coarse_lattice_linear_equation}
\end{equation}
From this solution on the coarsened lattice, one can construct
an approximate solution on the original fine lattice:
$|x\rangle_c \rightarrow |x\rangle$ (called `prolongation').
This is a basic strategy of the multi-grid preconditioning.
In addition, some operation called `smoothing' is applied
so as to improve the overlap of the approximate solution with
the solution on the fine lattice, which contains the high
frequency modes being not incorporated in the coarse solver. 
The smoother can be applied either before or after the coarse
solver, or even both before and after.

Eq.~(\ref{eq:coarse_lattice_linear_equation}) is easier to
solve than the original linear equation
due to its smaller rank, and to be solved not necessarily with
high accuracy.
In lattice QCD simulation, 
as quark mass decreases, 
the Dirac matrix
$D$ becomes dominated by the low frequency modes that is well
approximated by the coarse solution vector.

\paragraph{Construction of coarse grid operator.}
To construct a coarse vector, we prepare $N_{\rm nv}$ so-called null
space vectors that contain contribution from the low modes;
$|i \rangle$, $i=1,\dots, N_{\rm nv}$.
Since the quality of $|i \rangle$ governs the efficiency of the multi-grid
preconditioner,
we describe a way to generate them later in this section.
Having prepared $|i\rangle$, 
we divide the original lattice into $N_D$ domains
each composed of typically $4^4$ sites.
Each domain, labeled with a capital letter as $X$,
is mapped to one site on the coarse lattice.
Let us express the $i$-th null space vector whose nonzero components
are restricted to a domain $X$ as $|X, i\rangle$.
At this stage
we have totally $N_{\rm nv} N_D$ independent vectors.
We further double the number of vectors effectively by applying
projections in the 4-component `spin' space into two subspaces
labeled by $s=+,-$%
\footnote{
We apply projection matrices
$P_{\pm} = (1\pm \gamma_5)/2$ in the spin space, 
where $\gamma_5$ is a $4\times 4$ matrix
defined in Eq.~(\ref{eq:gamma5_hermiticity})
and satisfies $(\gamma_5)^2=1$.
Noting that the matrix $D$ satisfies $\gamma_5$-Hermiticity $(\gamma_5 D)^\dagger = (\gamma_5 D)$ and 
that $\gamma_5$ approximately maps an eigenmode of $\gamma_5 D$
with eigenvalue $\lambda$ to a one with $-\lambda$,
one can show that use of the projected spin basis is equivalent 
to a low rank approximation of the Hermitian matrix $(\gamma_5 D)$.
}.
The total number of independent vectors is thus $2 N_{\rm nv} N_D$ 
which is labeled by $I=\{s, i, X\}$.
Since these projections into domain $X$ and spin subspaces is local and 
almost trivial in the implementation, 
we need to keep only $N_{\rm nv}$ vectors $|i\rangle$.

Let us assume that vectors $2 N_{\rm nv} N_D$ vectors 
$\{|I\rangle\}$ are orthonormalized: $\langle I | J \rangle = \delta_{IJ}$.
Denoting the component of a coarse vector $|x\rangle_c$ as $x(I)$,
\begin{equation}
 x(I) = \langle I | x \rangle .
\label{eq:restriction}
\end{equation}
This operation reduces the degrees of freedom from
$12 L_x L_y L_z L_t$ to $2N_{\rm nv} N_D$.

Prolongation is performed by
\begin{equation}
 | x' \rangle = \sum_{I} x(I) | I \rangle .
\label{eq:prolongation}
\end{equation}
Note that $|x\rangle$ and $|x'\rangle$ are in general different.
The coarse Dirac fermion matrix $D_c(I,J)$ is represented as
\begin{equation}
 D_c = \langle I | D | J \rangle .
\label{eq:coarse_matrix_construction}
\end{equation}

\paragraph{Numerical steps.}
To summarize with a little more details, the multi-grid preconditioner
is composed of the following steps.
\begin{description}

\item[(0) Building of null space vectors (setup stage)] \ 

\begin{description}

\item[(0-a) Initial setup:]
We start with $N_{\rm nv}$ independent random vectors.
In general, multiplying an approximate $D^{-1}$ amplifies low frequency modes.
For this purpose, several iterations of arbitrary
iterative solver algorithm may suffice.
Instead of simply using $D$ on fine lattice, we employ the
Schwartz alternating procedure explained below.
After the vectors are orthonormalized to $\langle I | J \rangle = \delta_{IJ}$,
the coarse Dirac matrix $D_c$ is determined accordingly to
Eq.~(\ref{eq:coarse_matrix_construction}).

\item[(0-b) Adaptive improvement:]
Once the initial null space vectors and coarse matrix are prepared,
we use the multi-grid preconditioner itself (steps (1)--(2) bellow)
as an approximate $D^{-1}$ to 
further amplify the low frequency modes in the null space vectors.
The coarse matrix is simultaneously updated accordingly.
This improvement process may be repeated several times.
\end{description}
\end{description}

\noindent
Provided the null space vectors and coarse matrix, we employ the following
multi-grid preconditioner.
Each step is a Richardson refinement,
$|x\rangle + M_i (|b\rangle - D|x\rangle) \mapsto |x\rangle$
 with $M_i \simeq D^{-1}$ ($i=1,2$).
\begin{description}
\item[(1) Refinement on the coarse grid:]
   The approximate solver $M_1$ is made of the following three steps.
\begin{description}
\item[(1a) Restriction:]
For a given source vector for the preconditioner,
Eq.~(\ref{eq:restriction}) is applied to make a vector on
coarse lattice.

\item[(1b) Coarse matrix solver:]
The linear equation on coarse lattice
(\ref{eq:coarse_lattice_linear_equation}) is solved with
a standard way.
We employ the BiCGStab algorithm with an additional stabilization \cite{Sleijpen1995} in this work.
This is not necessarily performed with high precision.

\item[(1c) Prolongation:]
The solution vector of Eq.~(\ref{eq:coarse_lattice_linear_equation})
is prolonged to make a fine vector with Eq.~(\ref{eq:prolongation}).
The prolonged vector is used in the refinement of
approximate solution on the original lattice.

\end{description}

\item[(2) Smoother:]
We insert a smoother after the coarse grid refinement.
As $M_2$ in the Richardson process,
Schwartz Alternating Procedure (SAP) with fixed iterations
is used in this work for which details are described below.

\end{description}
Note that once the null space vectors are obtained in the step
(0a) and (0b), they are repeatedly used for different right hand side vectors
$|b\rangle$.

\paragraph{Schwartz Alternating Procedure (SAP).}
Schwartz alternating procedure helps to reduce the communication
in solving large linear system.
It was introduced in \cite{Luscher:2003vf,Luscher:2003qa}
as a domain-decomposed preconditioner to the lattice QCD,
and then used as an efficient smoother in the multi-grid
algorithm \cite{Frommer:2013fsa}.

One first divides the lattice into subdomains
each containing 
{\it e.g.} $4^4$ sites which should be practically determined
by observing the numerical efficiency.
One defines the Dirac matrix $D_{\rm SAP}$ by turning off the
interaction across the domains.
This implies that if each domain lies within a node,
inter-node communication does not exist in $D_{\rm SAP}$.
We set the SAP domain to be the same as the domain of coarsening
in the multi-grid algorithm.
Splitting the domains into even and odd groups, $D_{\rm SAP}^{(e)}$
and $D_{\rm SAP}^{(o)}$,
practical Schwartz alternating procedure is defined as follows.
\bigskip\\
\begin{tabular}{l}
\hline
\noalign{\smallskip}
Schwartz Alternating Procedure (SAP) with fixed iterations\\
\noalign{\smallskip}
\hline
\noalign{\smallskip}
$ |x\rangle := 0, \hspace{0.2cm}  |r\rangle := |b\rangle$ \\
for $i$ = 1, \dots, $n$ \ do \\
\hspace{0.5cm} $|x\rangle := |x\rangle + M_{\rm SAP}^{(e)}|r\rangle$ \\
\hspace{0.5cm} $|r\rangle := |b\rangle - D|x\rangle$ \\
\hspace{0.5cm} $|x\rangle := |x\rangle + M_{\rm SAP}^{(o)}|r\rangle$ \\
\hspace{0.5cm} $|r\rangle := |b\rangle - D|x\rangle$ \\
end \\
\noalign{\smallskip}
\hline
\end{tabular}
\bigskip \\
In this algorithm $M_{\rm SAP}^{(e/o)}$ is an inner-domain solver,
$M_{\rm SAP}^{(e/o)} \simeq (D_{\rm SAP}^{(e/o)})^{-1}$.
For this approximate solver, we adopt the MINRES algorithm with fixed
number of iterations, 6.
While the number of SAP iterations $n$ is one of the tuning
parameters in the multi-grid algorithm, we set it to 4 in this work.

\paragraph{Related works.}
Applications of the Multi-grid algorithms to lattice QCD have 
been developed in last 15 years \cite{Brannick:2007ue,Babich:2010qb,Osborn:2010mb}.
Not only for the Wilson-type fermion matrix applied in this work,
it is applied to the domain-wall fermion \cite{Cohen:2012sh}.
The SAP is introduced to the multi-grid algorithm in
Ref.~\cite{Frommer:2013fsa}.
Their implementation was ported to K-computer \cite{Ishikawa:2018fad}.
As for the optimization for recent architecture, application to
the Xeon Phi series is investigated \cite{Georg:2017zua}.
Among widely used lattice QCD code sets,
Grid \cite{grid, Boyle:2016lbp} contains a branch with multi-grid solvers.
QUDA \cite{quda, Clark:2009wm}, which is a lattice QCD library on GPUs,
also
contains multi-grid solvers.
The data layout for Fugaku in the current work is a generalization 
of a solver library (not multi-grid) for Fugaku \cite{QWSlibrary}.

\section{Implementation}
\label{sec:Implementation}

\subsection{Multi-grid algorithm}

To apply the multi-grid algorithm to variety of
computer architecture, it is desirable to separate the
algorithm and code implementation on specific architecture
as much as possible.
A guideline to accomplish it is the so-called object-oriented
programming.
As a code framework, we employ a general purpose lattice QCD
code set Bridge++ \cite{Bridge++, Ueda:2014rya} which is written in C++
based on the object-oriented design.
Bridge++ has been used to investigate a recipe of tuning on
Intel AVX-512 architectures \cite{Kanamori:2017a,Kanamori:2018}.

\begin{figure}[t]
\centering
\includegraphics[width=10.0cm]{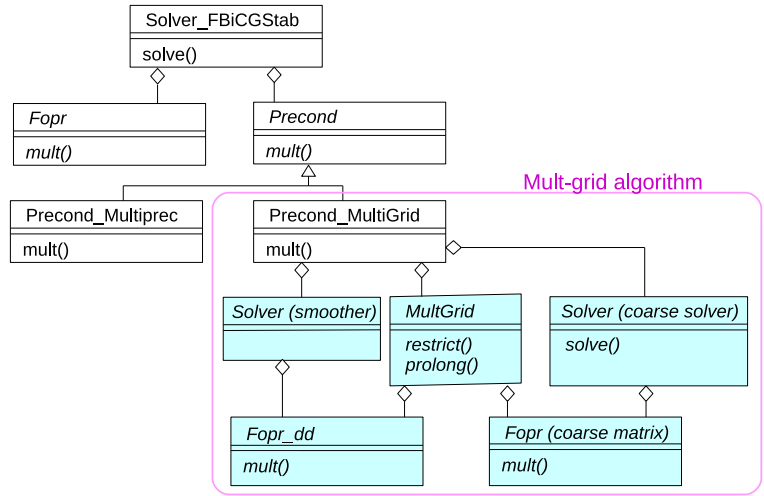}
\caption{
Class diagram for the multi-grid algorithm.
The objects of classes in blue color are implemented in
single precision arithmetics.
}
\label{fig:class_diagram}
\end{figure}

Figure~\ref{fig:class_diagram} displays a simplified class diagram
of our implementation.
All the classes are implemented as C++ template class
for which a template parameter {\tt AField} is omitted in the Figure.
{\tt AField} itself is a template class that holds
field data with specific precision and architecture
such as {\tt AField<float, SIMD>},
where the second template parameter is an {\tt enum}
entry.
The objects in the blue colored classes are 
constructed with single precision arithmetics.
We show only the abstract classes except for the top level
solver (BiCGStab algorithm with flexible preconditioning)
and two subclasses of the preconditioner, {\tt Precond}.
One may choose a multi-precision solver as a preconditioner
instead of the multi-grid algorithm.

Most of the ingredients of multi-grid algorithm is
implemented independently of the fermion matrix and architecture.
For a specific fermion matrix, {\it e.g.} the clover fermion in this work,
one needs to implement subclasses of {\tt Fopr} and {\tt Fopr\_dd},
where the latter represents a domain-decomposed version of the former.
One can apply optimization for the specific architecture to this implementation.
While the smoother is represented as an abstract {\tt Solver} class
and can be any solver, we use SAP solver in this work.
In a subclass of {\tt MultiGrid} class, functions for
restriction and prolongation are implemented for the specific
fermion matrix.
A blocked version of linear algebra is used in these operations.
Such blocked linear algebraic functions may effect the performance
significantly and thus are implemented
as a functional template code for each architecture.

In the case of the clover fermion, we therefore need to
implement three objects in addition to the standard fermion
matrix: the domain-decomposed version of the fermion matrix
({\tt AFopr\_Clover\_dd} class), the matrix on the coarse lattice
({\tt AFopr\_Clover\_coarse} class), and a set of blocked version
of linear algebraic functions such as {\tt dotc} and {\tt axpy}.
The last ingredients are provided as C++ template functions.
These parts are possible to optimize independently from
the construction of the algorithm.
In the following, we summarize the features of our target
architectures and procedures of tuning.

\subsection{Implementation for Intel AVX-512 architecture}

\begin{figure}[t]
\includegraphics[width=0.48\linewidth]{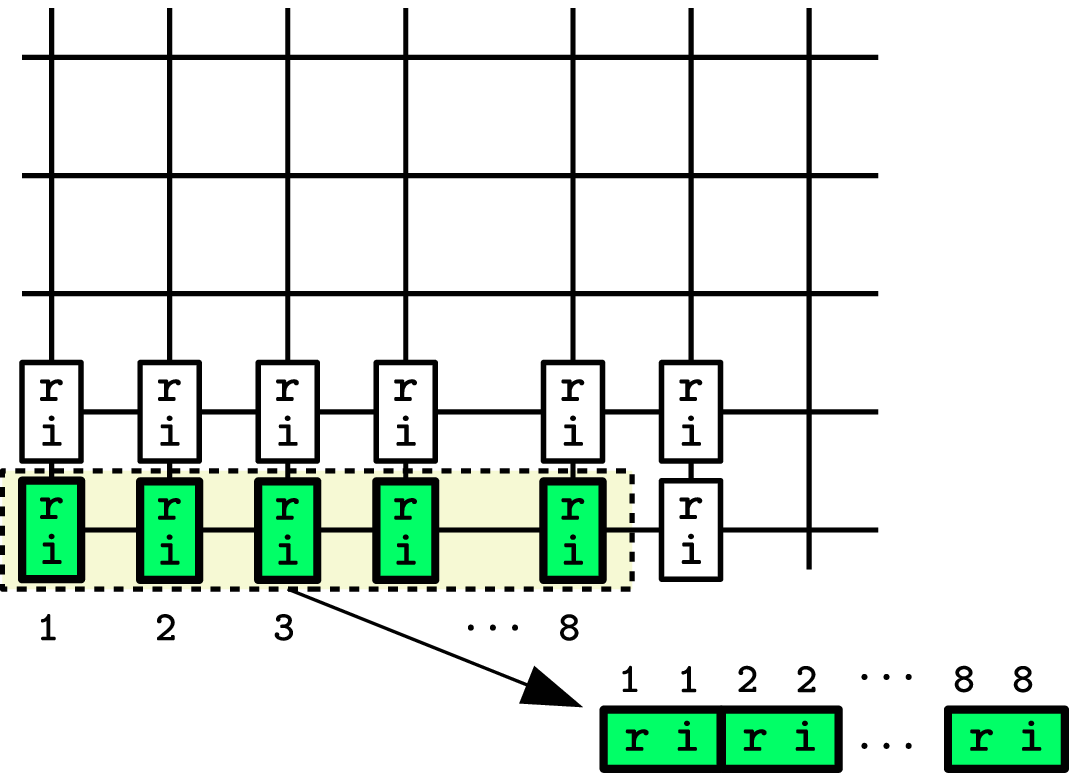}
\centering
\raisebox{0.2cm}{\includegraphics[width=0.48\linewidth]{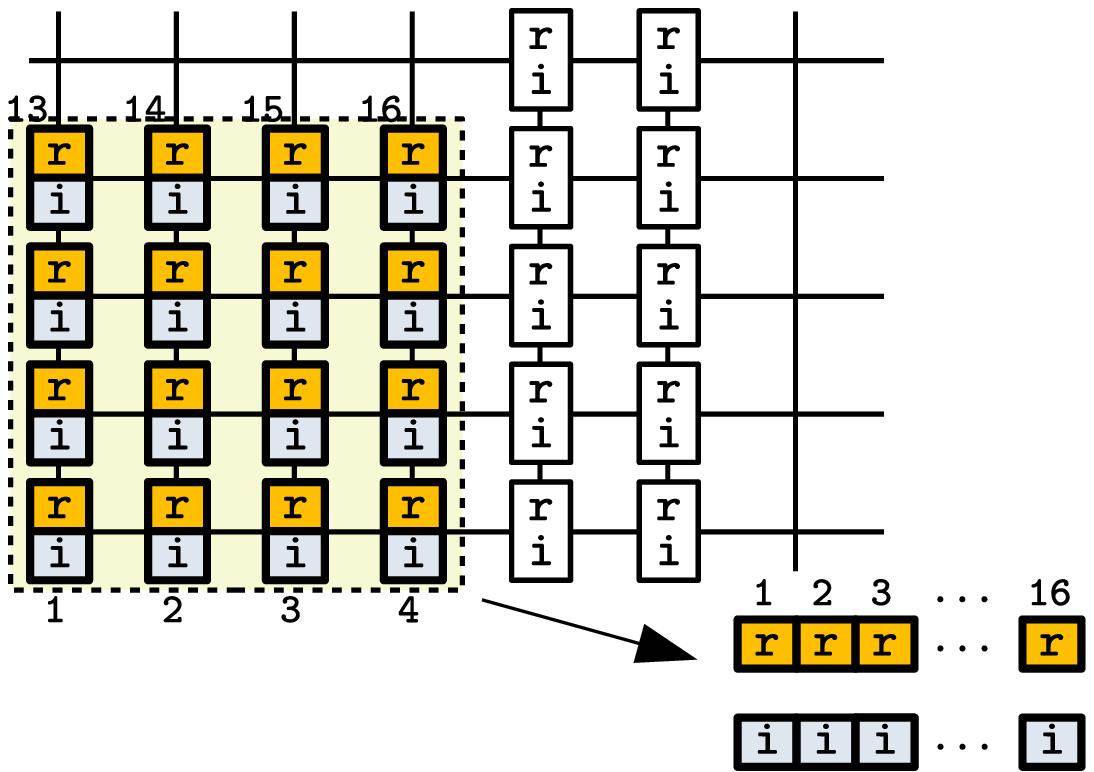}}
\caption{
The SIMD layouts for the AVX-512 (left panel)
and A64FX (right) architectures.  In the figure, \texttt{r} and \texttt{i} represent real and imaginary part of a complex number on each lattice site, respectively.  On the AVX-512, 8 complex numbers are packed to a SIMD variable while
16 complex numbers are packed to 2 SIMD variables (one for the real part and the other is the imaginary part) on the A64FX.
}
\label{fig:SIMD_layouts}
\end{figure}

We start with the implementation of our multi-grid code
for the Intel AVX-512 architecture.
Intel AVX-512 is the latest SIMD extension of x86 instruction set
architecture.
The 512-bit SIMD length corresponds to 8 double or 16 single
precision floating point numbers.
In the following we consider the single precision case with
which the preconditioner is implemented.
We pack 8 complex numbers that are consecutive in the
$x$-direction%
\footnote{This corresponds to the case (a) in
\cite{Kanamori:2017a,Kanamori:2018}.}
as displayed in the left panel of Figure~\ref{fig:SIMD_layouts}.
This implies that the size of the domain in $x$-direction
is a multiple of 8 and it is indeed fixed to 8 in this work.
For the coarse operator, we instead pack the internal degree of freedom to
the SIMD variables to avoid a strong constraint on the local coarse lattice size.
Since each coarse grid vector has $2N_{\rm nv}$ complex components,
$N_{\rm nv}$ is constrained to be a multiple of 4.

The implementation of the fine grid operators inherits 
the previous works \cite{Kanamori:2017a,Kanamori:2018}
and uses the L2 prefetch in the full operator $D$.
Although it has been tuned for $D$, 
the same prefetching pattern is used in the SAP operator.

\subsection{Implementation for A64FX architecture}

After the success of K computer, RIKEN decided to
develop the post-K computer as a massively parallel
computer with the Arm instruction set architecture with
scalable vector extension (SVE).
The development has successfully resulted in the Fugaku
supercomputer manufactured by Fujitsu that has been
installed at RIKEN R-CCS and provided for shared use
since March 2021.
In addition to Fugaku, there are several systems with
the same architecture.
The A64FX architecture adopted in Fugaku has 512-bit
SIMD length that amounts to 16 single precision floating
point numbers.
The SIMD instructions are directly specified in a C/C++
code through ACLE (Arm C Language Extension).

The linear equation solver for the Dirac matrix in
lattice QCD simulation has been one of the target
applications in the development of the Post-K project
that has lead to the Fugaku supercomputer.
As a product of so-called co-design study, an optimized code
named QWS (QCD Wide SIMD) library has been developed
\cite{QWSlibrary}.
Through this investigation, it has been found that better
performance is achieved by treating the real and imaginary
parts in different SIMD vectors.
Simplest implementation is achieved by packing 16 sites
in $x$-direction into one SIMD vector for FP32, as adopted
in the QWS library.
For the implementation of the multi-grid algorithm, however,
this implies that the domain size would be multiple of
16 at least in $x$-direction.
This is rather strong restriction for examining the
efficiency of the algorithm.
Thus we decided to develop another code that packs
the sites in $x$-$y$ plane as depicted
in the right panel of Figure~\ref{fig:SIMD_layouts}.
This enables for example to pack $4\times 4$ sites
into a single SIMD vector for FP32
and increases the flexibility in choosing the parameters.

We develop a library of fermion matrices with the above
2D SIMD packing and with the same convention and data layout
as QWS.
Tuning of this library is still underway by employing ACLE
by making use of the techniques established in the development
of QWS.

\subsection{Implementation for GPU clusters}

In implementation of a code for GPU, the first question is
which offloading scheme is to be adopted.
Two types of approaches are available: API-based libraries
(CUDA, OpenCL, etc.) and directive-based ones (OpenACC and OpenMP).
The former enables detailed manipulation of threads as well
as the use of local store shared by a set of device cores.
The latter is easy to start and suitable for incremental
development.
After preparatory study by comparing OpenCL and OpenACC,
Bridge++ adopted to develop the offloading code mainly using
OpenACC because of its simplicity and less effort in maintaining
the code.

For the fermion matrix on original lattice, we assign
the operations on each site to one device thread.
The data layout is changed so that the so-called
coalesced memory access is realized on the devices.
For the matrix-vector multiplication on coarse lattice,
computation of the vector component corresponding
to one null space vector on each coarse site is assigned
to one thread.
In the case of GPU clusters, severe bottleneck is data transfer
between the host processors and devices.
Thus we implement a code that minimizes such data transfer
and if necessary replace the general code with optimized code
by exploiting the specialization of template in C++.

\section{Performance result}
\label{sec:Performance}

We measure the performance of the multi-grid algorithm
on the following three lattice sizes of gauge configurations
labeled as A--C%
\footnote{
Although the condition numbers are not provided,
B has a significantly smaller value than the others.
The lighter pion mass implies the larger condition number,
which amounts 156, 512, and 145 MeV for the configuration
A, B, and C, respectively.
}.
\begin{itemize}
\item A: $32^3 \times 64$ lattice provided by PACS-CS collaboration
 \cite{Aoki:2008sm}.

\item B: $64^3 \times 64$ lattice provided by Yamazaki {\it et al.}
 \cite{Yamazaki:2012hi}.

\item C: $96^3 \times 96$ lattice provided by PACS collaboration
 \cite{Ishikawa:2015rho}.
\end{itemize}
These configurations are available through the Japan
Lattice Data Grid \cite{JLDG, Amagasa:2015zwb}.

The block size for the multi-grid setup is fixed to $8\times 4^3$.
The number of test vectors is set to 32 except for some benchmarks
of matrix multiplications.  Since these numbers should be tuned for 
each configurations, the throughput we investigate in this work
may not be optimal.
In addition to the solver, the performance of the
matrix-vector multiplications is examined in the weak scaling setting,
of which smaller local volume on Oakforest-PACS and Fugaku corresponds
to running set A on 16 nodes.

For comparison, we also measure the elapsed time to solve 
the same system with a mixed precision BiCGStab solver.
It uses the same flexible BiCGStab algorithm and the preconditioner
is a single precision stabilized BiCGStab \cite{Sleijpen1995},
the same algorithm used to solve the coarse system in the multi-grid solver.

\begin{table}[t]
\caption{
Elapsed time for the multi-grid solver.
The `solve' is for solving excluding the setup time.
The ratios tabulated are time for solving the coarse solver,
running the smoother, restriction and prolongation (R/P),
and the other in the multi-grid preconditioner.
For comparison, timing of mixed precision BiCGStab
is also tabulated as `MBiCGs'.
The block size is fixed to $8\times 4 \times 4 \times 4$.
`OFP' denotes Oakforest-PACS.}
\label{tab:etime_solvers}
\centering
 \begin{tabular}{ccccccccccc}
\hline
\noalign{\smallskip}
  sys. & conf.  & \# of & $N_{\mathrm{nv}}$ & \multicolumn{2}{c}{elapsed time [sec]} & \multicolumn{4}{c}{fraction in the solve.} & MBiCGs \\
       & label & nodes &                   & setup & solve & coarse & smoother & R/P  & prec.\ other & [sec] \\
\noalign{\smallskip}
  \hline
\noalign{\smallskip}
  OFP
  & A & 16 & 16 &   5.6 &  35 & 29\%  & 53\% & 5.2\% &  7.6\% & 54 \\
  & A & 16 & 32 &   16  &  23 & 62\%  & 26\% & 4.5\% &  4.0\% &      \\
  & B & 16 & 32 &    161 &  28 & 27\%  & 50\% &  13\% &  5.9\% &  41 \\
  & B & 32 & 32 &     65 &  13 & 28\%  & 50\% &  11\% &  6.1\% &  22 \\
  & C & 16 & 32 &    813 & 763 & 44\%  & 35\% &  11\% &  4.8\% & 2801\\
  & C & 96 & 32 &    113 & 175 & 53\%  & 29\% & 6.5\% &  5.5\% & 499 \\
  & C & 192 & 32&     60 &  80 & 51\%  & 28\% & 5.6\% &  5.9\% & 297 \\
\noalign{\smallskip}
\hline
\noalign{\smallskip}
  Fugaku
& A &  16 & 32 &   107 &  31 & 33\% & 40\% & 19\% & 3.2\% &  121\\
& B &  16 & 32 &   893 &  51 & 17\% & 48\% & 23\% & 6.6\% &   97 \\
& C &  96 & 32 &   688 & 213 & 34\% & 42\% & 18\% & 3.1\% & 1125 \\
& C & 216 & 32 &   314 & 111 & 41\% & 34\% & 18\% & 2.8\% &  512 \\
\noalign{\smallskip}
\hline
\noalign{\smallskip}
 Cygnus
& A &  1 & 32 & 500 &   198 & 50\% & 26\% & 20\% & 2.7\% & --- \\
& A & 16 & 32 &  78 &    79 & 37\% & 13\% & 47\% & 1.7\% & --- \\
& B & 16 & 32 & 263 &    21 & 23\% & 40\% & 27\% & 4.8\% & --- \\
\noalign{\smallskip}
\hline
 \end{tabular}
\end{table}

\begin{figure}[t]
\noindent
 \centering
\hspace{-0.2cm}
\includegraphics[width=0.51\linewidth]{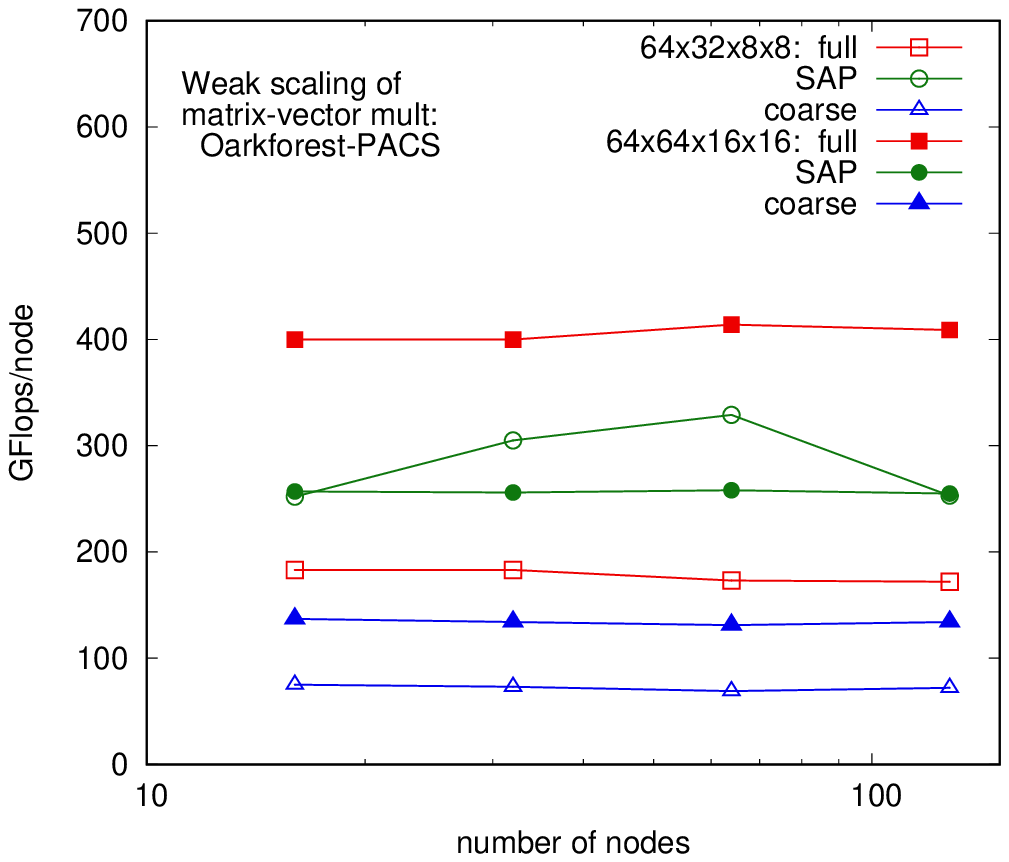}
\hspace{-0.4cm}
\includegraphics[width=0.51\linewidth]{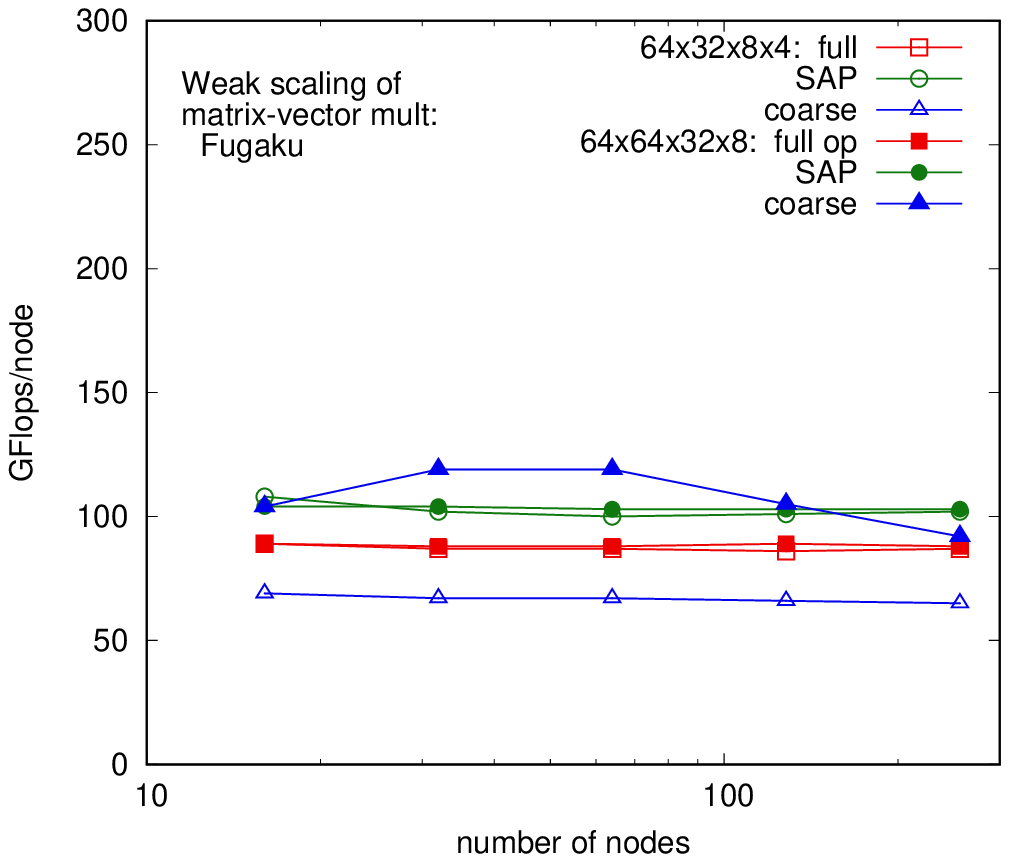}
\caption{
Weak scaling of matrix-vector multiplication on Oakforest-PACS (left panel)
and Fugaku (right panel) with two different local lattice volumes,
$5.2 \times 10^5$ and $4.2 \times 10^6$ lattice points per node.
The matrices are full matrix $D$, domain decomposed operator
$D_{\rm SAP}$ used in SAP in the smoother, and the coarse operator
used in the coarse solver.
The block size is $8\times 4 \times 4 \times 4$ and number of
null space vector $N_{\rm nv}$ is 32. }
\label{fig:performance_matvec_ofp_fugaku} 
\end{figure}

\subsection{Performance on  Intel Xeon Phi cluster}

We use the Oakforest-PACS system of JCAHPC.
The compiler is Intel compiler 2019.5 and the option used is
 \verb|-O3 -no-prec-div -axMIC-AVX512|. 
We use cache mode of the MCDRAM and set the number of the thread
to 2 per core.

In the left panel of Figure~\ref{fig:performance_matvec_ofp_fugaku},
the weak scaling of the matrix-vector multiplication is plotted in
two cases of local volume.
For the smaller local volume case,
the domain-decomposed operator is faster than the full operator
as expected from the absence of communications.
The tuning described in \cite{Kanamori:2018} for full operator works
quite efficiently for a larger local volume so that
its performance exceeds that of
the domain-decomposed operator.
Note that a naive roof line limit of the full operator is about 450 GFlops.
The coarse operator is slower than the others.
This is because the local volume is smaller than the fine lattice
so that the cost of
the neighboring communication becomes relatively larger,
in addition to the lack of detailed tuning such as prefetching.

The elapsed time and time budget are listed in Table~\ref{tab:etime_solvers}.
Once the null space vectors are prepared (`setup'), the multi-grid 
algorithm is faster than the mixed-precision solver.
However, the overhead of setup becomes not negligible as the condition
number decreases, as exhibited for configuration B.
The fraction of the coarse solver and smoother depends on the 
configuration and the number of nodes, while the remaining
part of the preconditioner
including the restriction/prolongation is kept small.
The performance of the multi-grid solver is 120--190 GFlops per node,
which depends on the size of local volume and the fraction of
the coarse grid solver.

\subsection{Performance on Fugaku}

On Fugaku, we use the Fujitsu compiler with the Clang mode.
The version of the compiler is 1.2.31 and the option is \verb|-Nclang -fopenmp -Ofast|.
When we measured the performance, May 2021, jobs with a small number of nodes ($<385$) are not guaranteed to have physically close allocation in the
6-dimensional mesh-torus network.
In this setting, frequent neighboring communication of
the matrix multiplication kernel of QCD may interfere with communication of other jobs.
We therefore secure a large enough number of nodes for which continuous torus
geometry is guaranteed and then use a subset of them
by specifying a rank map.

The performance of each matrix-vector multiplication is plotted
in the right panel of Figure~\ref{fig:performance_matvec_ofp_fugaku}.
There is plenty of room for improvement, since the node of Fugaku
has roughly the same peak performance as Oakforest-PACS
and larger memory bandwidth.
Note the difference of the scale between panels in the figure.
The SAP operator is fastest than the full operator as expected.
Although the code is still in the middle of tuning,
the coarse operator with larger local volume sometimes 
outperforms the SAP operator. 
All kernels show a good weak scaling.

The elapsed time and time budget are listed in Table~\ref{tab:etime_solvers}.
Compared with Oakforest-PACS, restriction and prolongation take
larger fractions.
This is because we have applied ACLE tuning to only a part of matrix
vector multiplication, and the linear algebraic functions
have still not been tuned.
The solving time is shorter than the mixed precision BiCGStab solver
in all cases.
Taking the setup time into account, the latter is faster for configurations
A and B.
Note that once the setup is finished, one can solve the linear equation
with different right hand sides repeatedly, which is a typical situation
in lattice QCD simulation.
In such a case, the multi-grid solver may become an efficient solution.
The total performance of the solver is 83 -- 96 GFlops per node, which
is in between the performance of coarse matrix and $D_{\rm SAP}$.

\subsection{Performance on GPU cluster}

As an example of GPU cluster, we use the Cygnus system
at University of Tsukuba.
Each node of Cygnus is composed of two Intel Xeon processors
(Xeon Gold 6126, 12 cores/2.6 GHz) and four NVIDIA Tesla V100 GPUs.
Although 32 nodes of Cygnus have FPGA devices connected
with dedicated network, we do not use this type of nodes.
Each V100 GPU has 5120 CUDA cores which amounts to
14 TFlops for FP32 arithmetics.
The memory bandwidth of the GPU global memory is 900 GB/s.
The bus interface is PCIe 3.0 with 16 lanes.
The nodes are connected by Infiniband HDR100 x4 network.
We use the NVIDIA HPC SDK 20.9 compiler with CUDA 11.0 and
the MPI library MVAPICH2-GDR 2.3.5.
The code is compiled with options \texttt{-O2 -ta=tesla:ptxinfo,cc70}.
We found that {\tt -fast} option rather decreases
the performance of the device kernel.
Since most arithmetically heavy operations are performed
on GPUs, multi-threading on the host processor contributes
the performance little,
and thus we switched it off
in this measurement.

\begin{figure}[t]
\centering
\includegraphics[width=0.65\linewidth]{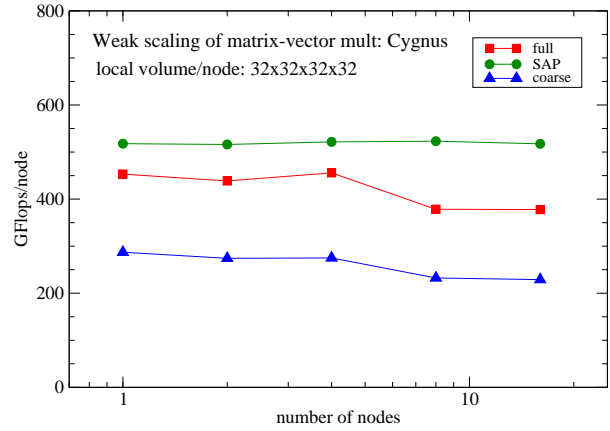}
\caption{
Weak scaling plot of the performance of matrix-vector
multiplications on Cygnus.
The local volume per node is set to $32^4$.
}
\label{fig:performance_matvec_cygnus}
\end{figure}

Figure~\ref{fig:performance_matvec_cygnus} displays the
weak scaling plot of the performance of matrix-vector
multiplications for
$D$ (full), $D_{\rm SAP}$, and coarse matrix $D_c$
in single precision.
To keep large enough local volume even after the
coarsening, we measure the performance with
lattice size per node $32^4$.
This lattice is divided to four GPU devices on each node.
We observe that the performances of $D$ and
coarse matrix decrease about 15\% between 4
and 8 nodes while that of $D_{\rm SAP}$ stays unchanged,
as the effect of changing the MPI parallelization
in $(x,y)$ to $(x,y,z)$ directions.
Considering the device memory bandwidth above and
the byte-per-flop of the matrix $D$, 0.94,
further optimization may be possible.
For this purpose, CUDA or OpenCL would be
required to employ for more detailed tuning.

The elapsed time and time budget are listed in
Table~\ref{tab:etime_solvers}.
The configuration C is not available due to insufficient
memory size of the maximum number of nodes (16) in our budget.
On Cygnus, we do not measure the mixed-precision
solver due to the lack of time for preparation.
General tendency of 
the time budget in the multi-grid algorithm is similar to
the SIMD architectures.
The fractions of restriction and prolongation, however,
tend to be larger due to
the data transfer between the hosts and devices
required for these steps in the current implementation.
The result implies that the current implementation is also
efficient for the architecture with GPU accelerators.

\section{Conclusion}
\label{sec:Conclusion}

In this paper, we implemented a multi-grid solver for lattice
QCD on two SIMD architectures and GPU clusters.
On all the architectures, the multi-grid solver accomplished
sufficient acceleration of elapsed time in solving linear
equation under typical parameter setup.
Although setup of the null space vector requires
non-negligible time, once they are prepared, solving the
linear equation for each source vector is significantly
accelerated.
In the measurement of physical quantities in lattice QCD
simulations, one frequently encounters the situation to
repeat such processes, for which the multi-grid algorithm
would be a powerful solution.

By separating the architecture-specific code from the
construction of algorithm, it has become possible to
tune for each architecture independently.
While present implementation is not very optimized,
one can concentrate on each architecture for individual
requirement.
In this paper, we focus on one type of fermion matrix.
Application to other type of matrix, such as the domain-wall
fermion, is underway.
Application to other recent architectures, such as the NEC
SX-Aurora TSUBASA with vector architecture, is also
planned.

\section*{Acknowledgment}
The authors would like to thank the members of lattice QCD
working group in FS2020 project for post-K computer and
the members of Bridge++ project for valuable discussion.
Numerical simulations were performed on Oakforest-PACS and
Cygnus systems through Multidisciplinary Cooperative Research
Program in CCS, University of Tsukuba, and supercomputer Fugaku
at RIKEN Center for computational Science.
Some part of code development were performed on the supercomputer
`Flow' at Information Technology Center, Nagoya University, and
Yukawa Institute Computer Facility.
We thank the Japan Lattice Data Grid team for providing the
public lattice QCD gauge ensemble through the grid file system.
This work is supported by JSPS KAKENHI (Grant Numbers 20K03961 (I.K.),
19K03837 (H.M.)),
the MEXT as ``Program for Promoting Researches on the Supercomputer
Fugaku'' (Simulation for basic science: from fundamental laws of
particles to creation of nuclei).
and Joint Institute for Computational Fundamental Science (JICFuS).

\end{document}